# Phonons in honeycomb and auxetic two-dimensional lattices

**A. Sparavigna**

**Dipartimento di Fisica**
**Politecnico di Torino**
**C.so Duca degli Abruzzi 24, 10129 Torino, Italy**

The modes of vibrations in honeycomb and auxetic structures are studied, with models in which the lattice is represented by a planar network where sites are connected by strings and rigid rods. The auxetic network is obtained modifying a model proposed by Evans et al. in 1991, and used to explain the negative Poisson's ratio of auxetic materials. This relevant property means that the materials have a lateral extension, instead to shrink, when they are stretched. For what concerns the acoustic properties of these structures, they absorb noise and vibrations more efficiently than non-auxetic equivalents.

The acoustic and optical dispersions obtained in the case of the auxetic model are compared with the dispersions displayed by a conventional honeycomb network. It is possible to see that the phonon dispersions of the auxetic model possess a complete bandgap and that the Goldstone mode group velocity is strongly dependent on the direction of propagation. The presence of a complete bandgap can explain some experimental observations on the sound propagation properties of the auxetic materials.

**Introduction**
Auxetic material are characterised to have a Poisson's ratio that turns out to be negative [1]: this relevant property means that these materials have a lateral extension, instead to shrink, when they are stretched. These materials are more resistant to indentations that ordinary materials because at the site of impact they become more dense and therefore more resistant. Natural auxetic materials and structures occur in biological systems, in skin and bone tissues. Man made auxetic materials have been produced from the nanoscale till the micro- and macroscales and cover the major class of materials, such as metals, ceramics, polymers and composites [2].

Although a negative Poisson's ratio is not forbidden by thermodynamics, this property is usually believed to be rare in crystalline materials: in contrast to this believe, 69% of cubic elemental metals have a negative Poisson's ratio [3]. Foams with auxetic structures built with polymeric materials are instead well known [4] and actually produced by Scott Industrial [5,6]. These auxetic materials can be involved in the production of seals, gaskets, energy absorption components and sound-proofing materials. Filter and drug-release materials will be soon produced, inserting auxetic fibers in technical textiles [7]. Some auxetic liquid crystal polymers have been studied too [8].

An interesting family of materials, very attractive for future applications in biological systems, are the crystalline membranes, which are auxetics as recently shown by Bowick et al. [9]. Let us remember here just one example of natural crystalline membrane, the cytoskeleton of the red blood cells [10,11]. Another proposal for applications in biology is the use of expanded auxetic PTFE as an ideal material for arterial prostheses, because it approximates more closely to the behaviour of natural biomaterial than the non-auxetic synthetic currently used.

A two-dimensional model for an auxetic mechanical system is that proposed in Ref.[1] and shown in Fig.1, where the auxetic honeycomb structure is on the left in comparison with the conventional honeycomb lattice on the right. From now on, the auxetic honeycomb structure will be simply called "auxetic", and the conventional honeycomb simply "honeycomb". The auxetic model shown in Fig.1 was introduced to easily explain the behaviour of a material with a negative Poisson's ratio: the vibration properties of such models were not subject of investigations.

The auxetic structure can be viewed as a structure composed by a network with flexible fibers of length $L$ but also with rigid rods between sites, those with length $L'$. It is then a structure where nodes are forming a two-dimensional lattice. We can also consider the honeycomb lattice with rigid connections between sites: for instance, all the bonds parallel to one of the lattice directions can be considered as rigid rods. These models are very stimulating for investigations of vibration modes, to verify the possibility of complete bandgaps in the phonon spectrum and the existence of gapped modes. In fact these models, both honeycomb and auxetic, represent lattices where the breaking of a discrete symmetry is introduced and a preferred direction appears, that parallel to the rigid bonds $L'$.

Moreover, experiments show that auxetic structures absorb noise and vibrations more efficiently than non-auxetic equivalents [5,12]: an explanation for this behaviour can surely come from the behaviours of the low energy modes and from the existence of complete bandgaps in the vibration spectra.

Bandgaps are intervals of frequencies where no propagating phonons exist. They are easily found in crystalline lattices where optical branches are separated from acoustic branches for all the directions of phonon propagation in the lattice. Cubic SiC for example is one of these lattices, a slightly polar compound where the lattice has a basis with two atoms of different mass [13,14]. Three-dimensional periodic elastic media can show bandgaps [15], and for this reason they are called "phononic crystals", as the photonic crystals displaying bandgaps for light waves. With phononic crystals, wave guides can be obtained too [16]. Bandgaps and optical gapped modes with a frequency cut-off, that is with a lowest frequency under which phonon modes cannot propagate, are now extensively investigated to analyse thermal transport in phononic crystals and mesoscopic systems, where quantized thermal transport becomes relevant [17,18].

Recently [19], the phonon dispersions have been discussed for membrane-like lattices with square geometry and non-uniform mass distribution along the strings. The honeycomb membranes, with or without rigid rods inserted in the structure, were not investigated. When the honeycomb network is stretched, the strings have a tension $T_o$ and the membrane can oscillate in transversal modes, that is with sites oscillating perpendicularly to the plane of the network. In the case of the auxetic structure shown in Fig.1, the strings are subjected to tension $T_o$ when the network is compressed. But in this case, the structure has the equilibrium position of the lattice sites which is not stable. To have stability and oscillations perpendicular to the sheet plane, we have to modify the model in Fig.1.

Before discussing the stable auxetic model and the honeycomb structure, let us briefly study a one-dimensional model to explain the method for solving the problem with the Bogoliubov transformation. Then the two-dimensional systems with the honeycomb and auxetic structure, as networks composed of irreducible cell units containing two nodes, will be solved. The mechanical models contain flexible strings and rigid rods. The three-dimensional structures, honeycomb and auxetic, will be considered in another paper.

**One-dimensional model.**

The one dimensional model is a chain composed of rigid units and string connections, where $L'$ is the length of the rigid unit with mass $M$, and $L$ is the length of the string connecting two massive units (see Fig.2). The unit cell of the lattice has a position given by the vector $(L + L')$. For simplicity, let us consider $L = L'$. The positions of the lattice sites (0) are denoted by the lattice indices $i, i+1, i+2,...$, and the sites of the basis are denoted by $B$. The mass per unit length of the rigid rod is $\rho'$. Ropes have a linear density $\rho$. Due to the geometry, $T_o$ is the equilibrium axial force in each string line unit. The lattice elements, strings and rods, are attached to each other at the nodal points.

Let us investigate the harmonic vibrations of the chain supposed to be infinite with displacements of lines and nodes in a transversal direction. $u_{i,b}$ is the displacement of one of the two nodes in the basis of the lattice cell at the reticular position $i$ from the equilibrium. $b$ can have two possible determinations $0$ and $B$. With $w_{ji,B0}$, it is called the displacement of a string connecting a node $i$ in the lattice cell with the nearest neighbour node $j$. A linear co-ordinate $\zeta$ is ranging from 0 to the length $L_{ji,B0}$ of the string. For the strings, the equation in the case of transverse vibrations is the usual wave equation, with a phase velocity $v = \sqrt{T_o/\rho}$. Solving the equation of motion for the string line units, we have as in Ref.19:

$$w_{ji,B0}(\zeta) = u_{i,B}\cos(\kappa\zeta) + \frac{u_{j,0} - u_{i,B}\cos(\kappa L)}{\sin(\kappa L)}\sin(\kappa\zeta) \tag{1}$$

in the case of a time-harmonic oscillation of frequency $\omega$, and $\kappa = \omega/v$.

In the chain there are strings and rods: an easy approach is to consider the rod as a rigid body and write the equations for the motion of the centre of mass and for rotations around it of the rod in a plane perpendicular to the chain. In the case of small displacements of the lattice nodes, equations are:

$$\frac{d^2(u_{i,B} + u_{i,0})}{dt^2} = \frac{T_o}{M}\left(\frac{dw_{ij,B0}}{d\zeta} + \frac{dw_{ij',B0}}{d\zeta}\right) \tag{2}$$

$$\frac{d^2(u_{i,B} - u_{i,0})}{dt^2} = \frac{L^2 T_o}{2I}\left(\frac{dw_{ij,B0}}{d\zeta} - \frac{dw_{ij',B0}}{d\zeta}\right) - \frac{LT_o}{2I}(u_{i,B} - u_{i,0}) \tag{3}$$

where index $j$ denotes the lattice site connected with the site $i$ on the right, and index $j'$ the site connected with $i$ on the left of the rigid unit. $dw_{ij,B0}/d\zeta, dw_{ij',B0}/d\zeta$ are proportional to the force components perpendicular to the chain and then acting on the rigid body. The force component parallel to the chain is simply considered equal to $T_o$. If we are looking for Bloch waves with wavevector k, it is possible to write for each lattice site:

$$u_{i,0} = u_0 \exp(-2ikL) \quad ; \quad u_{i,B} = u_B \exp(-2ikL) \tag{4}$$

and then the dispersion relations for the frequency $\omega$ can be easily obtained from the dynamical equations (2) and (3) of the rods. To simplify the solution we use a Bogoliubov rotation of $u_0, u_B$. The Bolgoliubov transformation is the following:

$$\begin{aligned} u_0 &= \eta + i\theta \\ u_B &= \eta - i\theta \end{aligned} \tag{5}$$

and then we have:

$$-M\omega^2\eta = \frac{T_o}{vS}\{\eta\omega(\cos(kL)-C)+\theta\omega\sin(kL)\}$$

$$-2\frac{I}{L^2}\omega^2\theta = \frac{T_o}{vS}\{-\theta\omega(\cos(kL)+C)-S\theta+\eta\omega\sin(kL)\}$$
(6)

where $C=\cos(\kappa L)$; $S=\sin(\kappa L)$. For $\sin(\kappa L)=0$, for any $\kappa$ standing wave, modes exist corresponding to internal vibration of the strings, with no associated modal displacements. Let us consider the reduced frequency $\Omega = \omega/\omega_o$ where $\omega_o = T_o/Mv$. The dispersion relations of the chain as a function of the wavenumber $k$ are shown in Fig.2 for different values of the ratio $I/ML^2$. It is interesting to note the existence of a gap in the phonon dispersion between acoustic and optical modes.

**Honeycomb and auxetic lattices.**

The first two-dimensional model here considered is a planar membrane with a honeycomb structure, as shown in Fig.1 on the right: it is possible to see the unit cell of the lattice with a convenient set of hexagonal vectors $(\mathbf{l}_1,\mathbf{l}_2)$ giving the lattice reticular positions. The lattice is simple hexagonal; the positions of the lattice points are denoted by black circles in the figure, with one extra node per unit cell, as shown in Fig.4. The sites in the basis are denoted by open circles. We can imagine the lattice made with rigid rods, with length $L$ and mass per unit length $\rho'$, and ropes with length $L$ and linear density $\rho$. In the figure, the thick lines are representing the rigid units. $T_o$ is the equilibrium axial force in each string line unit which is giving a tensile action on the rods of the honeycomb lattice.

Since the honeycomb network is represented by a hexagonal lattice, the first Brillouin Zone is hexagonal too. The directions OX, OC and OY used for calculations are shown in the Fig.4.

Let us investigate the harmonic vibrations of the honeycomb net supposed to be infinite with displacements of lines and nodes in the direction perpendicular to its plane. As in the case of the one-dimensional lattice, $u_{i,b}$ is the displacement of one of the two nodes in the basis of the lattice cell at the reticular position. The same for $w_{ji,B0}$, the displacement of a string connecting a node in the lattice cell with the nearest neighbour node. If we are looking for Bloch waves with wavevector $\mathbf{k}$, it is possible to write for each lattice cell the displacements as:

$$u_{i,0} = u_0 \exp(-i\mathbf{k}\cdot\mathbf{l}_1)$$
$$u_{i,B} = u_B \exp(-i\mathbf{k}\cdot\mathbf{l}_2)$$
(7)

The dispersion relations for the frequency $\omega$ can be easily obtained from the dynamics of the rods, solving the following equations:

$$-M\omega^2\eta = \frac{T_o}{vS}\{\eta\omega(\cos\mathbf{k}\cdot\mathbf{l}_1+\cos\mathbf{k}\cdot\mathbf{l}_2-C)+\theta\omega(\sin\mathbf{k}\cdot\mathbf{l}_1+\sin\mathbf{k}\cdot\mathbf{l}_2)\}$$

$$-2\frac{I}{L^2}\omega^2\theta = \frac{T_o}{vS}\left\{-\theta\omega(\cos\mathbf{k}\cdot\mathbf{l}_1+\cos\mathbf{k}\cdot\mathbf{l}_2+C)-S\theta\cos\frac{\pi}{3}+\eta\omega(\sin\mathbf{k}\cdot\mathbf{l}_1+\sin\mathbf{k}\cdot\mathbf{l}_2)\right\}$$
(8)

where $C = \cos(\kappa L)$ and $S = \sin(\kappa L)$. As in the one-dimensional chain, the Bogoliubov transformation is used to solve the dynamics. The reduced frequency $\Omega = \omega/\omega_o$ (where $\omega_o = T_o/Mv$) of the honeycomb two-dimensional lattice can then be evaluated for different values of the ratio $I/ML^2$ as a function of the wavevector $\mathbf{k}$, in the directions OX, OC and OY, where O is the centre of the Brillouin Zone.

In the case of the auxetic lattice, the rods have a length $L'$ different from the length $L$ of the ropes. Due to the geometry, $T_o$ is the equilibrium force in each string line unit which is giving a compression on the rods. As previously told, the auxetic network shown in Fig.1 must be modified to obtain a stable equilibrium configuration of the lattice. A simple solution is to insert other ropes binding the rigid rods. Let us suppose these strings with a tension $\xi T_o$. Comparing Fig.1 and Fig.4, it is easy to identify the ropes inserted for the stability. These ropes have a sound speed $v_2 = \sqrt{\xi T_o/\rho}$, different from the speed $v_1 = \sqrt{T_o/\rho}$ of the other ropes. As a result of all the forces acting on the rigid units, we have the following equations:

$$-M\frac{F}{T_o}\omega^2\eta = \eta\omega\left(c_1+c_2-C+\xi\frac{F}{F'}(c_3-C')\right)+\theta\omega\left(s_1+s_2+\xi\frac{F}{F'}s_3\right)$$

$$-\frac{I}{L'^2}\frac{F}{T_o}\omega^2\theta =$$

$$= -\theta\omega\left(c_1+c_2+C+\xi\frac{F}{F'}(c_3+C')\right)+\frac{F}{L'}\theta\cos\frac{\pi}{3}-\xi\frac{F}{L'}\theta+\eta\omega\left(s_1+s_2+\xi\frac{F}{F'}s_3\right)$$
(9)

where $C = \cos(\kappa_1 L)$; $S = \sin(\kappa_1 L)$; $C' = \cos(\kappa_2 L)$; $S' = \sin(\kappa_2 L)$; $F = v_1 S$; $F' = v_2 S'$; $C' = \cos(\kappa_2 L)$ and $\kappa_1 = \omega/v_1$, $\kappa_2 = \omega/v_2$. The coefficients containing the wavenumber are $c_1 = \cos \mathbf{k} \cdot \mathbf{l}_1$; $s_1 = \sin \mathbf{k} \cdot \mathbf{l}_1$; $c_2 = \cos \mathbf{k} \cdot \mathbf{l}_2$; $s_2 = \cos \mathbf{k} \cdot \mathbf{l}_2$; $c_3 = \cos \mathbf{k} \cdot (\mathbf{l}_1 + \mathbf{l}_2)$; $s_3 = \sin \mathbf{k} \cdot (\mathbf{l}_1 + \mathbf{l}_2)$.

In this case, the reduced frequency we use is $\Omega = \omega/\omega_o$, with $\omega_o = T_o/M v_1$. The Fig.5 shows the phonon dispersions of the auxetic (in red) lattice, for a value of ratio $I/ML'^2 = 1$. In green, the dispersions for the honeycomb lattice is shown for comparison ($I/ML^2 = 1$). Of course, the dispersion of acoustic modes is depending on the direction of propagation. When parameter $\xi$ is increased over a certain value ($\xi = 5$ for the $I/ML'^2 = 1$.), a complete bandgap between the acoustic and the optical mode appears. Let us reduce now the value of the ratio $I/ML'^2$: the parameter $\xi$ needs to be increased over a lower value (for instance, $\xi = 1.7$ for $I/ML'^2 = 0.1$) to observe the complete bandgap. Figure 6 shows the phonon dispersions for $\xi = 2$ and $I/ML'^2 = 0.1$. A further increase of this parameter gives a wider gap.

**Conclusions.**

We have seen that a complete bandgap is easy to obtain in two-dimensions, adjusting lattice parameters and interactions, that is changing elastic properties or densities of ropes and rods. Of course, different and more complex auxetic and honeycomb models must be proposed and studied, to exhaustively understand the behaviour of these structures. Moreover, we must create three-dimensional models to determine if complete bandgaps in all the bulk directions are possible. The study of auxetic three-dimensional lattices can then be very interesting for the development of phononic materials.

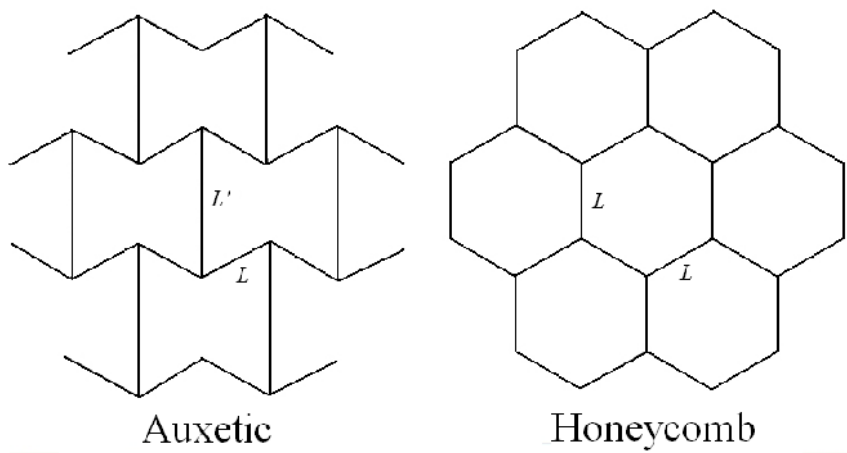

Fig.1: The auxetic structure compared with a honeycomb lattice. For the auxetic lattice the Poisson's ratio is negative: if the lattice on the left is horizontally stretched, it spreads in the vertical direction. $L$ and $L'$ are the lengths of strings and rods respectively.

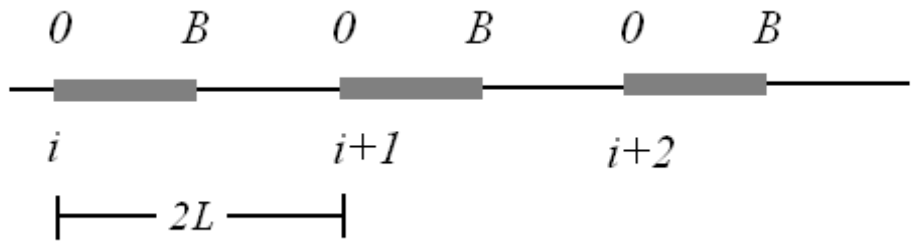

Fig.2: The chain with rigid units. $L$ is the length of the rods and strings.

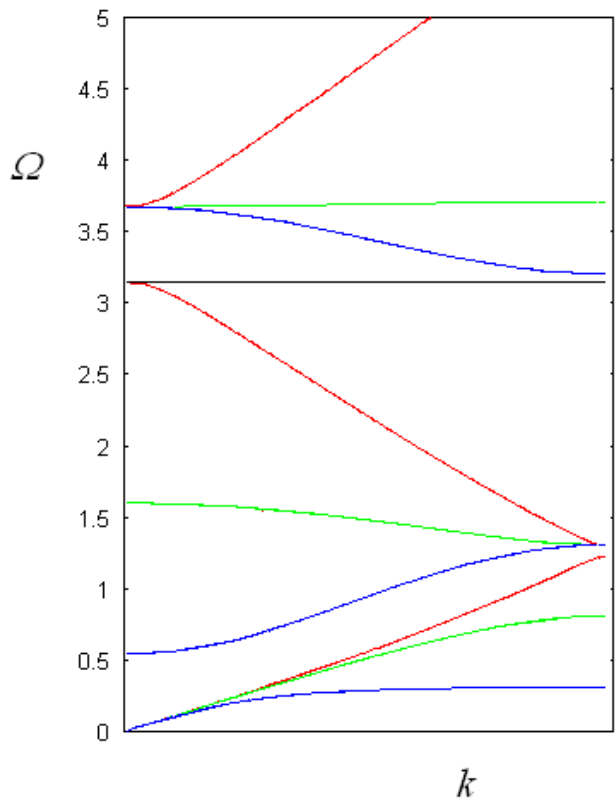

Fig.3: Dispersions of the chain for different values of the ratio $I/ML^2$ (0.1 in red, 1. in green and 10. in blue). Note the behaviour of the optical mode for high values of the ratio $I/ML^2$. The gap between acoustic and optical modes is quite pronounced for the green curve.

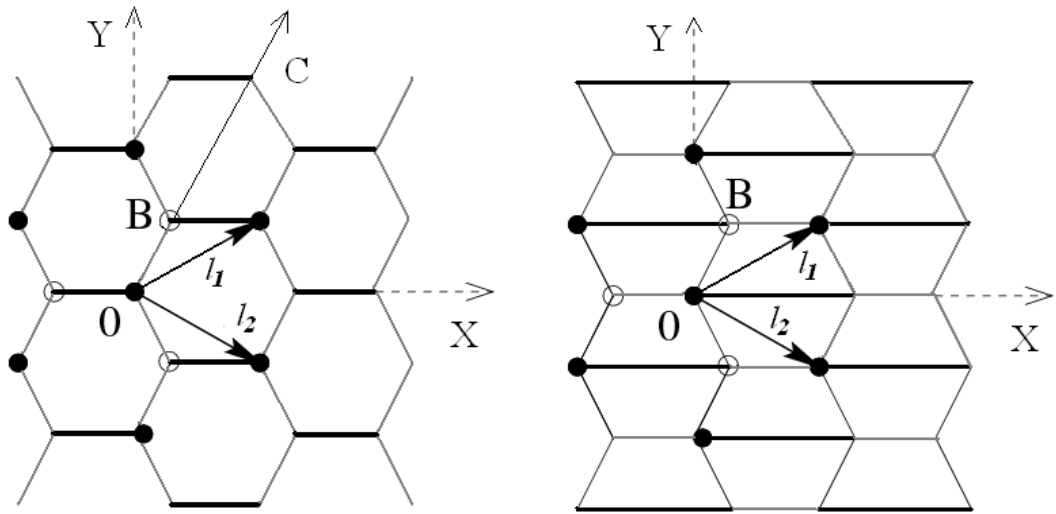

Fig. 4: The hexagonal lattice on the left, with the primitive lattice points (black dots) and the points of the basis (white dots), and a set of two lattice vectors convenient for calculations. Thick lines are representing the rigid rods. On the right, the auxetic model with additional strings for the stability of the lattice equilibrium position. The reciprocal lattice is hexagonal. The three directions OX, OC and OY along which the dispersion relations are evaluated are also displayed.

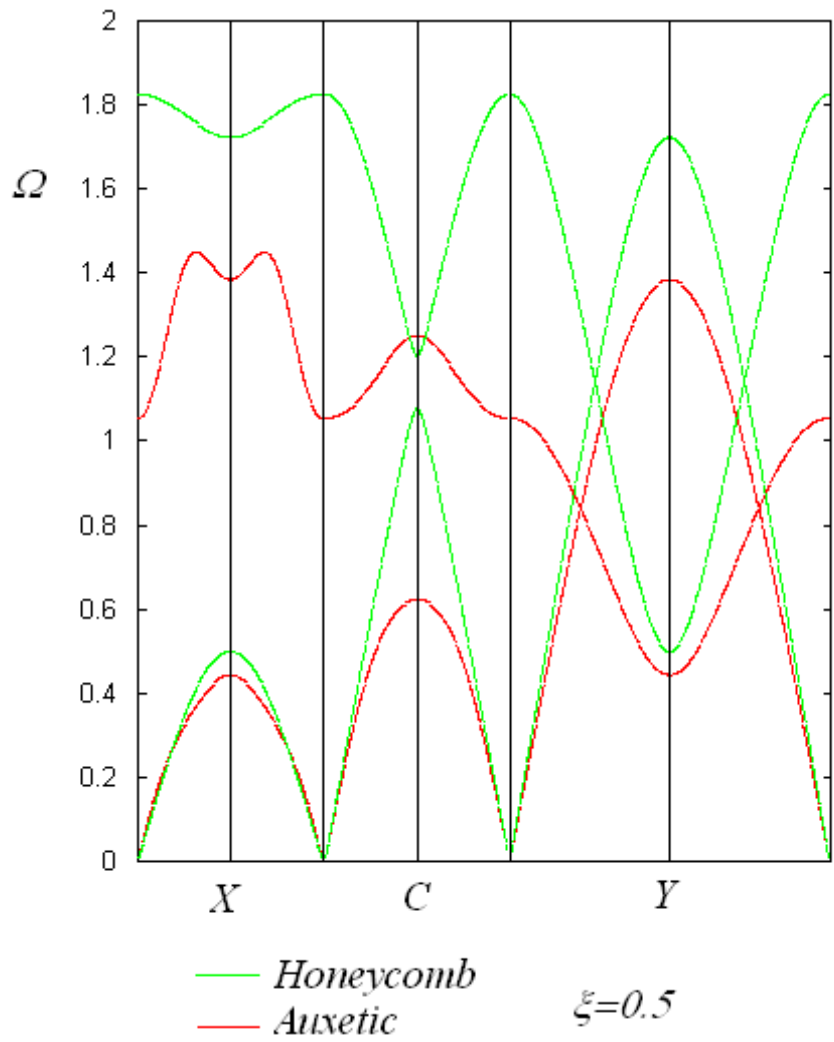

Fig.5: The phonon dispersions for the auxetic (green) and honeycomb (red) lattices for the ratio $I/ML^2 = 1$ in the case of $\xi = 0.5$.

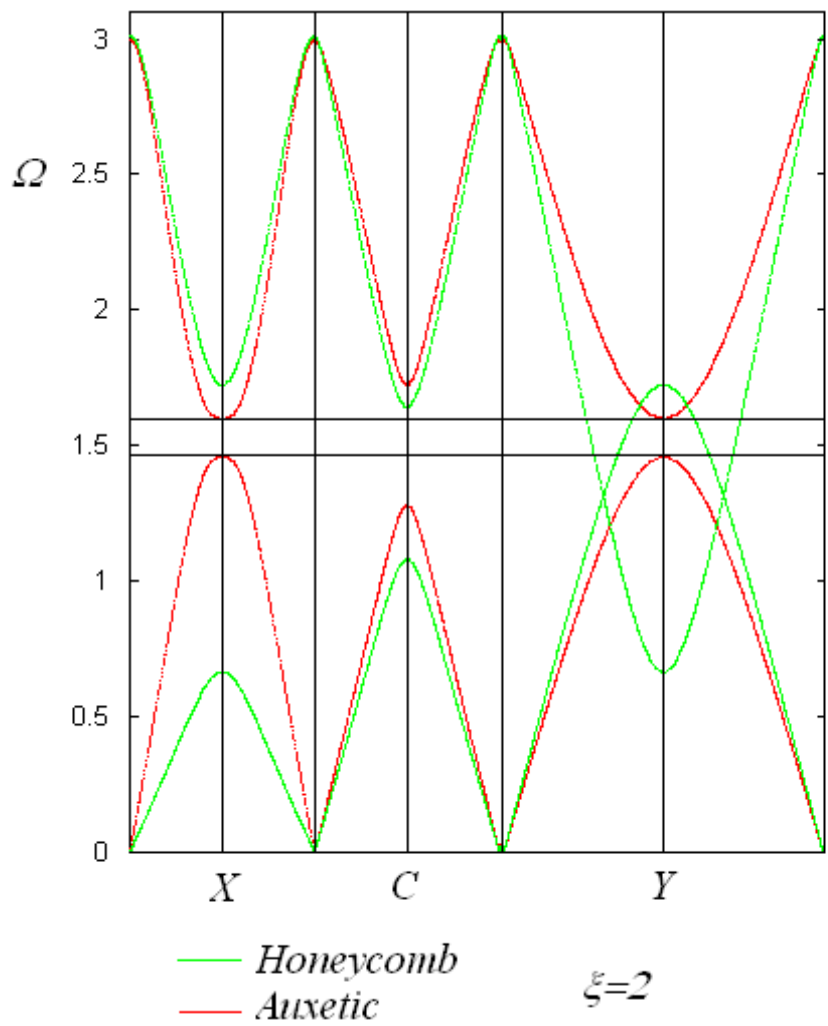

Fig.6: The same as in Figure 5 for $I/ML^2 = 0.1$ and $\xi = 2$. Note the complete bandgap of the auxetic structure.